\documentclass[12pt]{article}
\usepackage{lineno}
\usepackage{graphicx}
\usepackage{cite}
\usepackage{subcaption}

\newcommand\pubnumber{CMS experiment}
\newcommand\pubdate{\today}

\def\institute{Instituto de Ciencias y Tecnologías Espaciales de Asturias (ICTEA), Departamento de Física\\
Universidad de Oviedo}

\def\Title#1{\begin{center} {\Large #1 } \end{center}}
\def\Author#1{\begin{center}{ \sc #1} \end{center}}
\def\Address#1{\begin{center}{ \it #1} \end{center}}

\newcommand\pubblock{\rightline{\begin{tabular}{l} \pubnumber\\
         \pubdate  \end{tabular}}}
\newenvironment{Abstract}{\begin{quotation}  }{\end{quotation}}
\newenvironment{Presented}{\begin{quotation} \begin{center} 
             PRESENTED AT\end{center}\bigskip 
      \begin{center}\begin{large}}{\end{large}\end{center} \end{quotation}}

\def\beq{\begin{equation}}
\def\eeq#1{\label{#1}\end{equation}}
\def\eeqn{\end{equation}}

\def\beqa{\begin{eqnarray}}
\def\eeqa#1{\label{#1}\end{eqnarray}}
\def\eeqan{\end{eqnarray}}

\let\bar=\overbar

\def\Dslash{\not{\hbox{\kern-4pt $D$}}}
\def\dslash{\not{\hbox{\kern-2pt $\del$}}}

\def\msb{{\bar{\ssstyle M \kern -1pt S}}}

\def\qqbar{\mathrm{t}\bar{\mathrm{t}}}
\begin{document}
\begin{titlepage}
\pubblock

\vfill
\Title{Inclusive t${\bar{\mathrm{t}}}$  production cross section at $\sqrt{s}$ = 5.02 TeV with CMS}
\vfill
\Author{Carlos Vico Villalba on behalf of the CMS collaboration}
\Address{\institute}
\vfill
\begin{Abstract}
The top quark pair production cross section is measured in proton-proton collisions at a center-of-mass energy of 5.02 TeV. The data collected in 2017 by the CMS experiment at the LHC corresponding to an integrated luminosity of 304 pb$^{-1}$~are analyzed. The measurement is performed using events with one electron and one muon of  opposite  sign,  and  at  least  two  jets.   The  measured  cross  section  is  found  to  be $\sigma_{\mathrm{t}\bar{\mathrm{t}}}~=~60.3~\pm5.0~\mathrm{(stat)}\pm2.8~\mathrm{(syst)}\pm0.9~\mathrm{(lumi)}$~pb. To reduce the statistical uncertainty, a combination with the result in the $\ell$+jets channel, based on 27.4~pb$^{-1}$ of data collected in 2015 at the same center-of-mass energy of 5.02~TeV, is then performed, obtaining a value of $\sigma_{\mathrm{t}\bar{\mathrm{t}}}~=~62.6~\pm4.1~\mathrm{(stat)}\pm3.0~\mathrm{(syst+lumi)}$~pb, with a total uncertainty of 7.9\%, in agreement with the standard model.
\end{Abstract}
\vfill
\begin{Presented}
$14^\mathrm{th}$ International Workshop on Top Quark Physics\\
(videoconference), 13--17 September, 2021
\end{Presented}
\vfill
\end{titlepage}
\def\thefootnote{\fnsymbol{footnote}}
\setcounter{footnote}{0}

\section{Introduction}
The detailed study of the top quark constitutes one of the core elements of the CERN LHC physics programme. For instance, top quarks are mainly produced in top quark-antiquark ($\qqbar$) pairs due to gluon-gluon fusion being the most probable interaction between high energy protons. Therefore, $\qqbar$ production can be used to better determine the gluon parton distribution functions (PDFs) of the proton~\cite{CMS:2017iqf}. A variety of measurements have been performed by the different LHC experiments, in a wide range of decay channels and collision energies. The first measurement of the $\qqbar$ production cross section, in events with two jets and one or two leptons (electron or muon, $\ell$), using a data sample that corresponds to an integrated luminosity of 27.4~pb$^{-1}$ and at a center-of-mass energy of 5.02 TeV, was performed by the CMS experiment~\cite{CMS:2008xjf} in 2015~\cite{CMS:2010uxk}. 
Due to the reduced number of additional interactions per bunch crossing (pileup) with respect to the standard operating conditions of the LHC being one of the most appealing aspects of these data samples, a new run of data taking with a similar configuration was held in 2017. The recorded integrated luminosity of this new data sample amounts for 304~pb$^{-1}$ of considerably low pileup events. In this document, a measurement of the production cross section of $\qqbar$ ($\sigma_{\qqbar}$) in dileptonic events with two jets in the final state, which significantly improves the previous result, is presented. The final result is combined with the $\ell$ + jets result from \cite{CMS:2010uxk}.

\section{Analysis strategy}

Background contributions are modelled using several Monte Carlo (MC) generators such as \texttt{Madgraph\_5aMC@NLO}~\cite{Alwall:2014hca} and \texttt{Powheg} (v2)~\cite{Alioli:2010xd}, as well as for the evaluation of efficiencies and uncertainties. Interesting events are selected using the CMS trigger system, considering only events that fired at least one single lepton trigger. Since this analysis is based on leptonic selections, candidate events are required to have two leptons with opposite sign ($\mathrm{e}^{\pm}\mu^{\mp}$) and at least two jets. Due to the lepton identification being of great importance in these measurements, a multivariate discriminant, based on a gradient boosted decission tree~\cite{CMS:2020mpn}, is trained to separate between leptons that have been originated after a W or Z boson decay (known as prompt) and those who do not. Efficient triggering of leptons is ensured by requiring the leading lepton to have transverse momentum ($p_{\mathrm{T}}$) greater than 20 GeV. In addition, purity of signal over background in the final selection is enhanced by choosing an invariant mass threshold for the dilepton pair of 20 GeV, which significantly reduces the amount of background coming from photon conversions and low mass resonances.

\begin{figure}[htpb]
     \centering
     \begin{subfigure}[b]{0.32\textwidth}
         \centering
         \includegraphics[width=\textwidth]{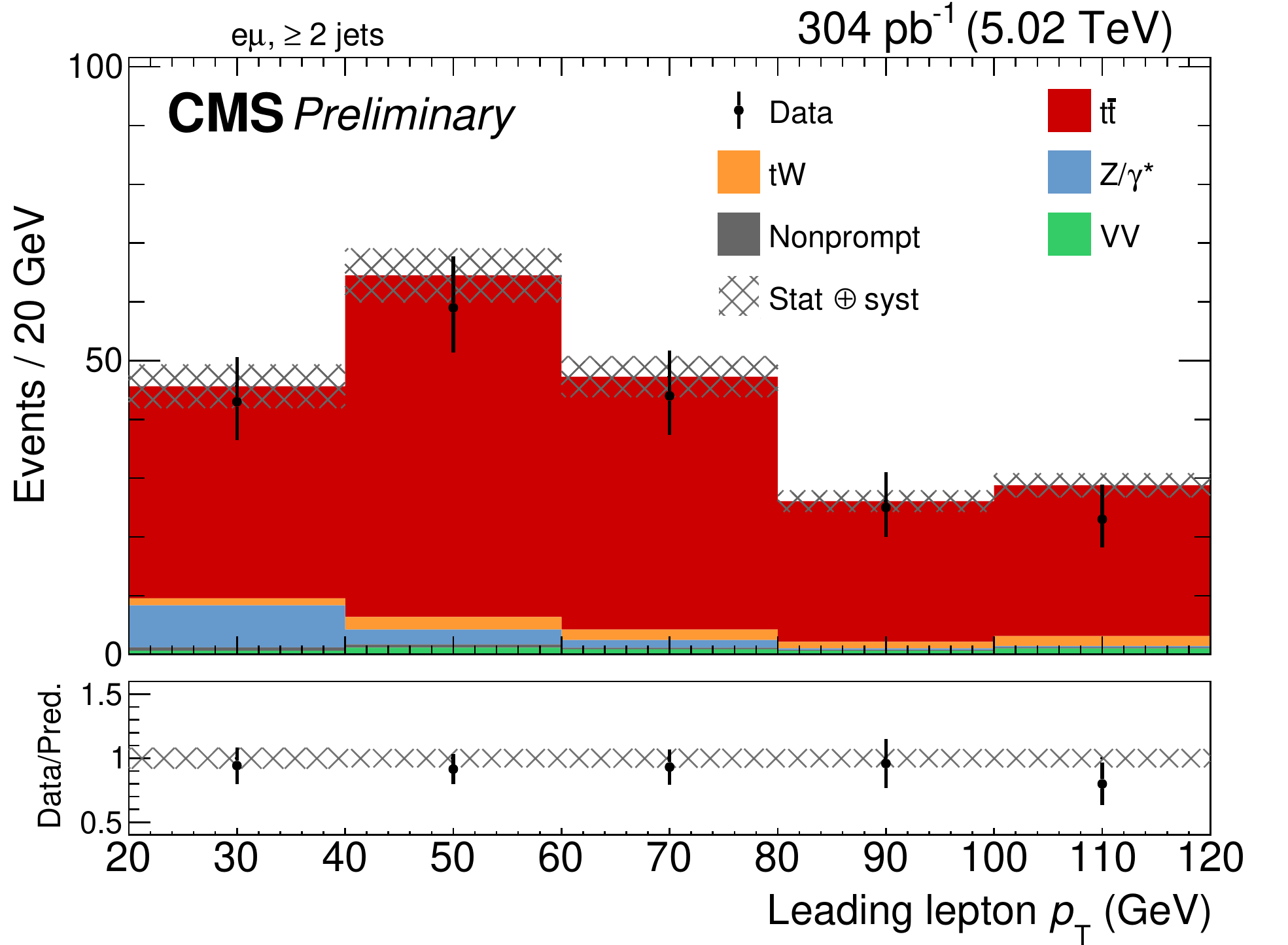}
         \caption{}
         \label{leadingLep}
     \end{subfigure}
     \hfill
     \begin{subfigure}[b]{0.32\textwidth}
         \centering
         \includegraphics[width=\textwidth]{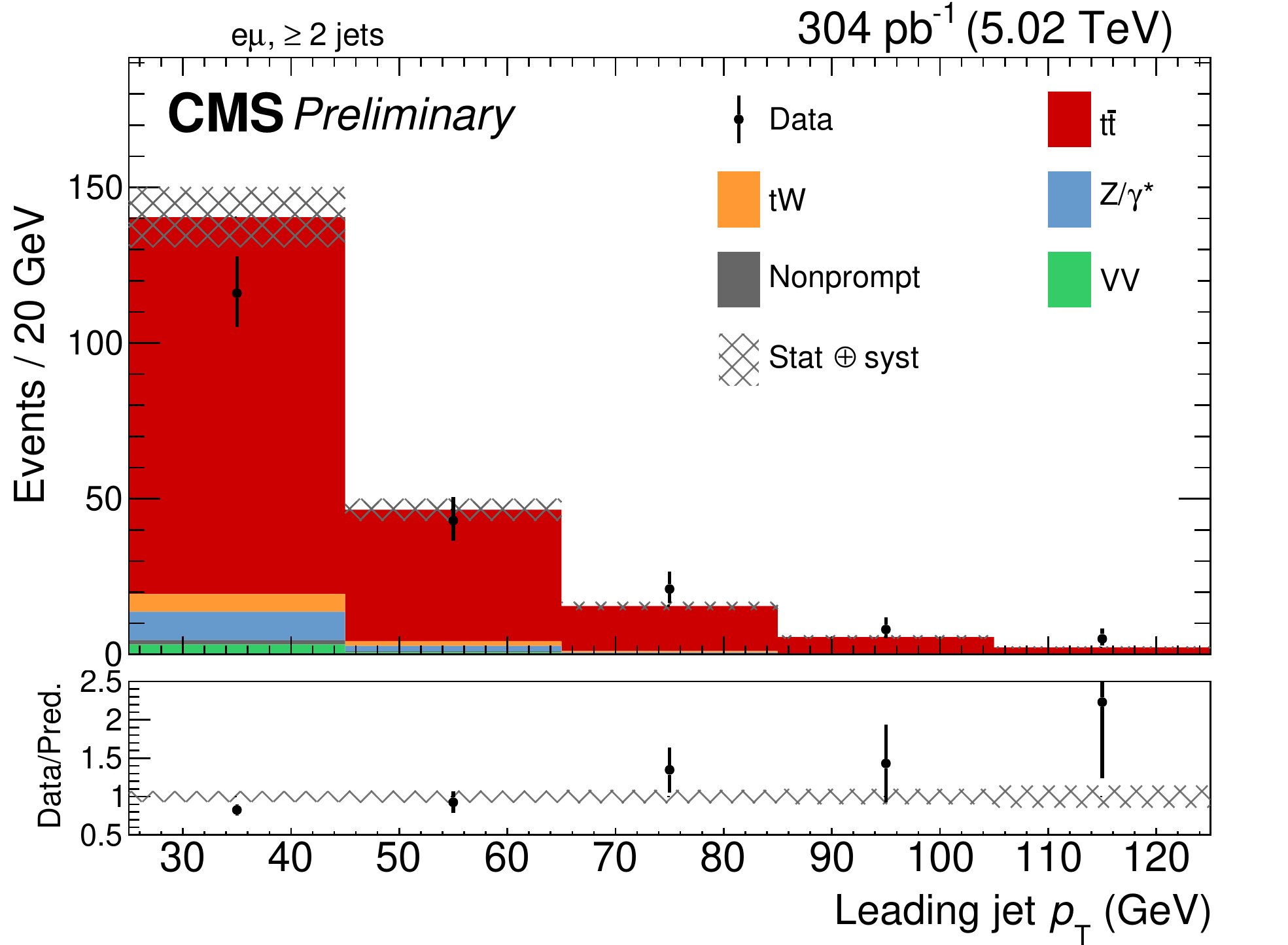}
         \caption{}
         \label{leadingJet}
     \end{subfigure}
     \hfill
     \begin{subfigure}[b]{0.32\textwidth}
         \centering
         \includegraphics[width=\textwidth]{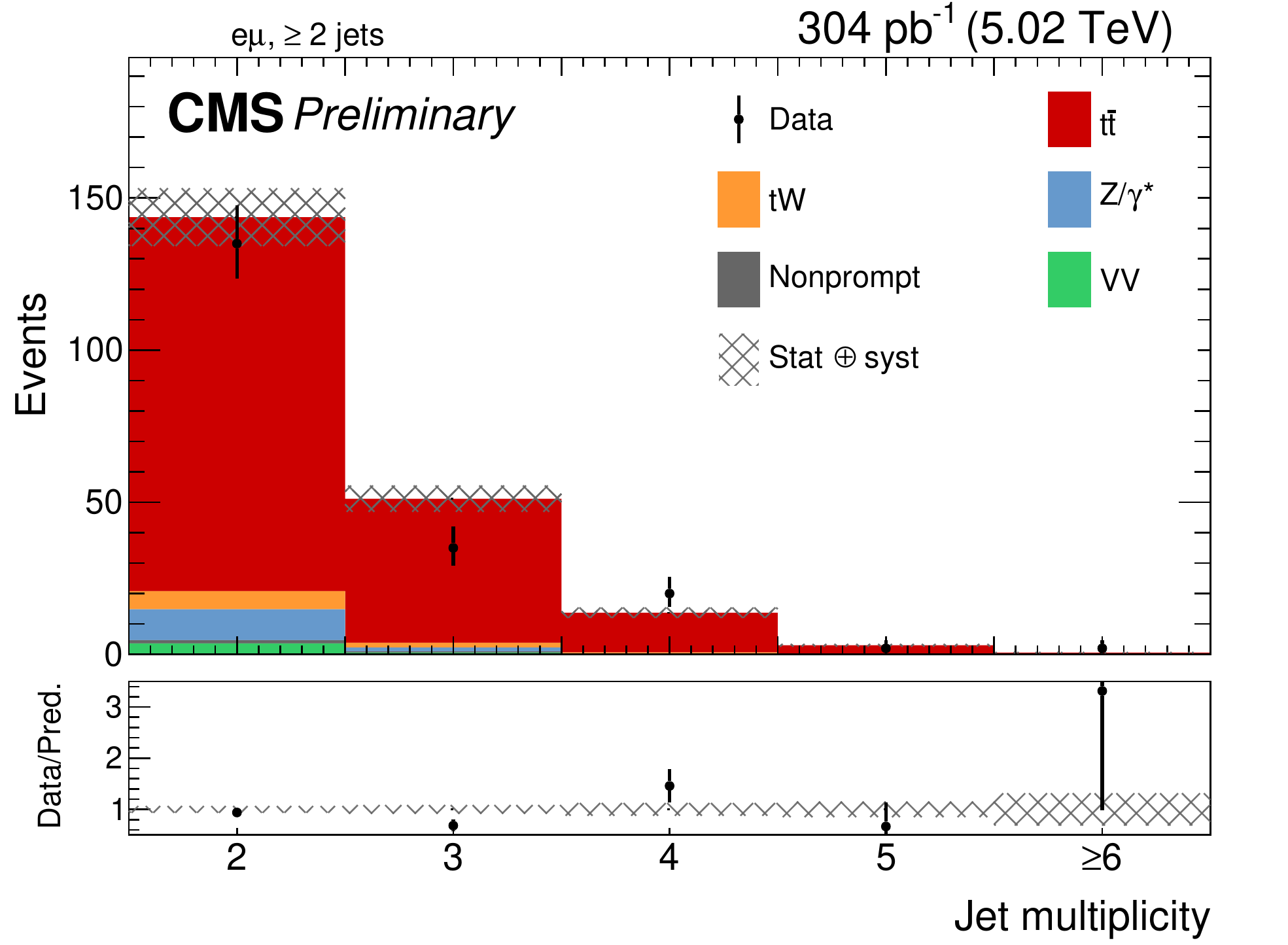}
         \caption{}
         \label{njet}
     \end{subfigure}
        \caption{Distributions of the leading lepton (left) and leading jet (center) $p_{\mathrm{T}}$, as well as the amount of background in the final state as a function of the number of jets (right). The total uncertainty band (shaded band in the plots) accounts for the systematic uncertainties, whereas the error bars account for the statistical uncertainty~\cite{CMS:2017zpm}.}
        \label{distributions}
\end{figure}

Figure~\ref{distributions} shows the $p_{\mathrm{T}}$ of the leading lepton (\ref{leadingLep}) and the leading jet (\ref{leadingJet}), and the background composition in non-inclusive regions that depend on the multiplicity of jets in the final state (\ref{njet}). Signal and background composition shows a considerably pure selection of $\qqbar$ events in the final state. The production of single top quarks in association with a W boson (tW), Drell-Yan (DY) and diboson (VV) production give rise to prompt leptons, and for that they constitute the main sources of background events for this analysis. While tW and VV background contributions are estimated using simulations only, the DY event yield is estimated in a data-driven way using the $R_{out/in}$ method, described in \cite{CMS:2010uxk}. Less contributing background sources such as semileptonic $\qqbar$ and W+jets, whose main presence in the final state comes from the production of non-prompt leptons, are grouped into the ``Nonprompt'' category. Reasonable agreement between data and prediction is also noticed in the lower panels placed below each figure.

\section{Systematic uncertainties}
 In Table~\ref{tab:ex}, a summary of the different sources of systematic and statistical uncertainty, and their effect in the final measurement of $\sigma_{\qqbar}$, is shown. The first four entries of the table account for theoretical uncertainties that arise from the normalization and estimation of the different backgrounds, and they have been evaluated by repeating the $\sigma_{\qqbar}$ extraction using dedicated simulation samples. The following sources account for uncertainties of experimental origin, and they have been calculated considering  $\pm1\sigma$ variations of the input parameters, so that an estimation on how the input to the fit affects the $\sigma_{\qqbar}$ extraction can be obtained. Lastly, the last three entries are the total systematic uncertainty (determined by adding the effect of all the individual systematic components in quadrature), the uncertainty related to the measurement of the integrated luminosity \cite{CMS:2021nvp}, and the statistical uncertainty. The result is dominated by the statistical uncertainty, while the uncertainty on the jet energy scale and the DY background estimate constitute the largest systematic uncertainties.
 
\begin{table}
    \centering
    \resizebox{0.4\textwidth}{!}{\begin{tabular}{lc}  
    Source &  $\Delta\sigma_{\qqbar}$/$\sigma_{\qqbar}$ (\%)\\ \hline
     tW  &   1.0 \\
     Nonprompt leptons &   0.4 \\ 
     Drell-Yan &   1.8 \\ 
     VV &   0.8 \\ \hline
     Trigger efficiency &  1.3 \\ 
     L1 prefiring &   1.4 \\ 
     Electron efficiency &   1.6 \\ 
     Muon efficiency &   0.6 \\ 
     JES &   2.2 \\ 
     JER &   1.2 \\ 
     $\mu_{R}$, $\mu_{F}$ scales &   0.2 \\ 
     PDF$\oplus\alpha_S(m_Z)$ &   0.3 \\ 
     Final state radiation &   1.1 \\ 
     Initial state radiation &   $<$ 0.1 \\ 
     h$_{\mathrm{damp}}$ &   1.0 \\ 
     Underlying event tune &   0.7 \\ \hline
     Total systematic &   4.3 \\ 
     Integrated luminosity &   1.5 \\ 
     Statistical uncertainty &   8.2 \\ 
    \end{tabular}}
    \caption{Summary of the statistical and systematic uncertainty sources on the $\qqbar$ cross section measurement~\cite{CMS:2017zpm}.}
    \label{tab:ex}
\end{table}

\section{Results}
The $\sigma_{\qqbar}$ measurement is performed through a counting experiment, following equation~\ref{eq:counting}. In this expression, N is the total number of observed events, N$_{\mathrm{bkg}}$ is the amount of background that remains after the selection of events, $\mathcal{L}$ is the integrated luminosity, $\mathcal{BR}$ is the branching ratio of $W\rightarrow e^{\pm}\mu^{\mp}$, used to account for the fact that the measurement is performed using only dileptonic final states; $\mathcal{A}$ is the total acceptance defined as the fraction of $\qqbar\rightarrow e^{\pm}\mu^{\mp}$ that meet selection requirements, and $\epsilon$ is the event selection efficiency.

\begin{equation}
    \sigma_{\qqbar} = \frac{N-N_{\mathrm{bkg}}}{\epsilon\cdot\mathcal{A}\cdot\mathcal{BR}\cdot\mathcal{L}}
    \label{eq:counting}
\end{equation}

\begin{figure}
    \centering
    \includegraphics[width = 0.55\textwidth]{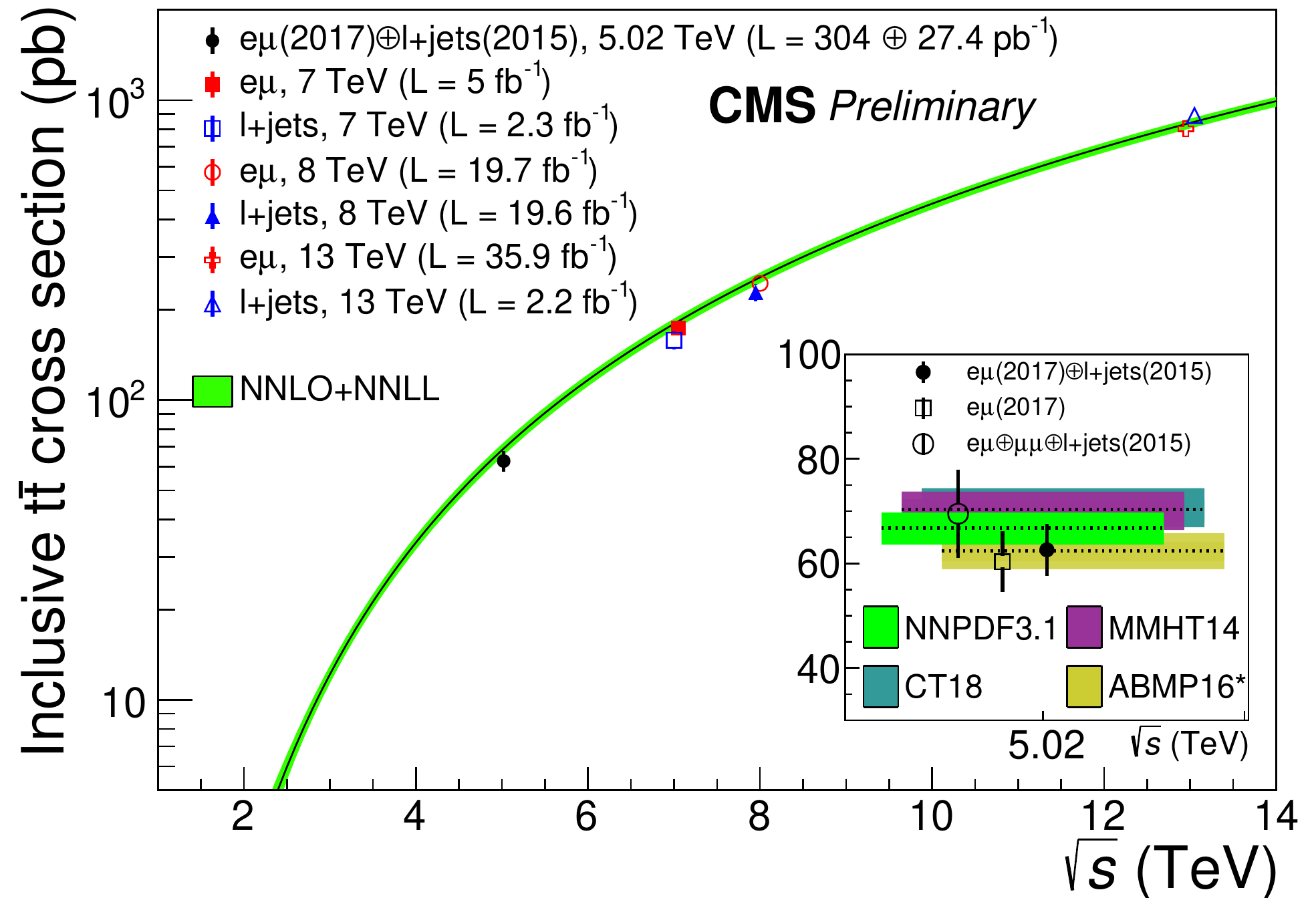}
    \caption{In the main plot, a summary of the different measurements of the inclusive $\qqbar$ cross section in pp collisions, at different center-of-mass energies, and their comparison to NNLO+NNLL theoretical predictions~\cite{Czakon:2013goa}; computed with the NNPDF3.1~\cite{NNPDF:2017mvq} PDF set with $\alpha_S$(m$_Z$) = 0.118 and $m_{\mathrm{top}}$=172.5~GeV. In the inset, the combined measurement presented in this analysis and its comparison with previous measurements and additional predictions at 5.02 TeV, computed with different NNPDF3.1 PDF sets \cite{Harland-Lang:2014zoa,Dulat:2015mca,Alekhin:2017kpj}. The vertical bars and bands represent the total uncertainties in the data and in the predictions, respectively~\cite{CMS:2017zpm}. 
    }
    \label{fig:summaryplot}
\end{figure}

The measured cross section is $\sigma_{\qqbar}~=~60.3~\pm~5.0~(\mathrm{stat})~\pm~2.8~(\mathrm{syst})~\pm~0.9~(\mathrm{lumi})~\mathrm{pb}.$ This measurement is statistically dominated due to the limited size of the data sample. In order to reduce this statistical limitation, the result has been combined with the one obtained in \cite{CMS:2010uxk}, considering that all sources of systematic uncertainty are uncorrelated between both analysis, since the data sets, $\qqbar$ modelling and background estimation methods are different; except for the normalization of the tW background and the QCD scale choice, which are fully correlated. The final result is  $\sigma_{\qqbar}~=~62.6~\pm~4.1~(\mathrm{stat})~\pm~3.0~(\mathrm{syst + lumi})~\mathrm{pb}$, which is found to be on agreement with the Standard Model prediction, as shown in Figure \ref{fig:summaryplot}, where a summary of CMS measurements of $\sigma_{\qqbar}$ at different energies are compared to the NNLO+NNLL prediction. In the inset, the measurement presented in this analysis is compared to different NNLO PDF sets.

\section{Summary}
A measurement of the top quark pair production cross section at a center-of-mass energy of 5.02 TeV is presented for events with one muon and one electron of opposite charge sign, and at least two jets, using proton-proton collisions collected by the CMS experiment in 2017 and corresponding to an integrated luminosity of 304 pb$^{-1}$. A measured cross section of  $\sigma_{\qqbar}$ =62.6$\pm$4.1~(stat)$\pm$3.0~(syst+lumi)~pb is obtained by combining the result from the analysis of the 2017 data set with the result presented in \cite{CMS:2010uxk}. The total uncertainty on this measurement is 7.9\%, which appart from being in agreement with the SM prediction, it provides an improvement of roughly 5.1\% with respect to the previous measurement.






\bibliography{eprint}{}
\bibliographystyle{unsrt}
 
\end{document}